\documentclass[sigconf,10pt]{acmart}

\usepackage{booktabs} 
\usepackage{listings}
\usepackage{amsmath}
\usepackage{algorithm}
\usepackage[noend]{algpseudocode}
\usepackage{balance}

\copyrightyear{2019} 
\acmYear{2019} 
\setcopyright{acmcopyright}
\acmConference[SIGMOD '19]{2019 International Conference on Management of Data}{June 30-July 5, 2019}{Amsterdam, Netherlands}
\acmBooktitle{2019 International Conference on Management of Data (SIGMOD '19), June 30-July 5, 2019, Amsterdam, Netherlands}
\acmPrice{15.00}
\acmDOI{10.1145/3299869.3314046}
\acmISBN{978-1-4503-5643-5/19/06}

\definecolor{dkgreen}{rgb}{0,0.6,0}
\definecolor{gray}{rgb}{0.5,0.5,0.5}
\definecolor{mauve}{rgb}{0.58,0,0.82}

\newcommand*{\affmark}[1][*]{\textsuperscript{#1}}

\begin{document}
       
\author{Haifeng Liu\affmark[\S \dag], Wei Ding\affmark[\dag], Yuan Chen\affmark[\dag], Weilong Guo\affmark[\dag], Shuoran Liu\affmark[\dag],\\Tianpeng Li\affmark[\dag], Mofei Zhang\affmark[\dag], Jianxing Zhao\affmark[\dag], Hongyin Zhu\affmark[\dag], Zhengyi Zhu\affmark[\dag]}

\affiliation{\affmark[\S]University of Science and Technology of China, Hefei, China}    

\affiliation{\affmark[\dag]JD.com, Beijing, China}
 

%
%
%
%
%
%
%
%
%

  \renewcommand{\shortauthors}{Haifeng Liu  et al.}



\title{CFS: A Distributed File System for Large Scale Container Platforms}

\begin{abstract}
We propose  CFS, a distributed file system for large scale container  platforms. CFS supports both sequential and random file accesses with  optimized  storage for both large files and small files, and adopts different replication protocols for different write scenarios to improve the replication performance.  It employs a metadata subsystem to store and distribute the file metadata across different storage nodes based on the memory usage. This metadata placement strategy avoids the need of data rebalancing during capacity expansion.  CFS also provides POSIX-compliant APIs with relaxed semantics and  metadata atomicity to improve the system performance.

We performed a comprehensive comparison with Ceph, a widely-used distributed file system on container platforms.  Our experimental  results show that, in testing 7 commonly used metadata operations,  CFS gives around 3 times  performance boost on average. In addition,  CFS exhibits better random-read/write performance in highly concurrent environments with multiple clients and processes.

\end{abstract}

\maketitle

{\fontsize{10pt}{10pt}\selectfont
\textbf{ACM Reference Format:}\\
Haifeng Liu, Wei Ding, Yuan Chen, Weilong Guo, Shuoran Liu, Tianpeng Li, Mofei Zhang, Jianxing Zhao, Hongyin Zhu, Zhengyi Zhu. 2019. CFS: A Distributed File System for Large Scale Container Platforms. In \textit{2019 International Conference on Management of Data (SIGMOD ’19), June 30-July 5, 2019, Amsterdam, Netherlands.} ACM, New York, NY, USA, 14 pages. https://doi.org/10.1145/3299869.3314046 }

\section{Introduction}
 
 
 



Containerization and microservices have revolutionized cloud environments and architectures over the past few years~\cite{bernstein2014containers, pahl2015containerization, balalaie2016microservices}. As applications can be built, deployed and managed faster through continuous delivery,  more and more companies start to move legacy applications and core business functions to containerized environment. 

The microservices running on each set of containers are usually independent from the local disk storage. While decoupling compute from storage allows the companies to scale the container resources in a more efficient way,  it also brings up the need of a separate storage because (1) containers may need to preserve the application data even after they are closed, (2) the same file may need to be accessed by different containers simultaneously, and (3) the storage resources may need to be shared by different services and applications.
Without the ability to persist data, containers might have limited usage in many workloads, especially in stateful applications.

One option is to take the existing distributed file systems and bring them to the cloud native environment  through the Container Storage Interface (CSI)\footnote{\url{https://github.com/container-storage-interface/spec}}, which has been supported by  various container orchestrators such as   Kubernetes~\cite{kubernetes} and Mesos~\cite{mesos}, 
 or through some storage orchestrator such as Rook\footnote{\url{https://rook.io/}}.  
 When seeking such a distributed file system, the  engineering teams who own the applications and services running on JD's container platform provide many valuable feedbacks. However, in terms of performance and scalability,  these feedbacks also give us hard time to adopt any existing open source solution directly.


For example, to reduce the storage cost, different applications and services usually need to be served from the same shared storage infrastructure. As a result, the size of files in the combined workloads can vary from a few kilobytes to hundreds of gigabytes, and these files can  be accessed in a sequential or random fashion. However, many distributed file systems  are  optimized for either large files such as HDFS~\cite{hdfs}, or small files such as Haystack~\cite{haystack}, but very few of them have optimized storage for both large and small size files~\cite{ceph, efs, azure, gluster}.  Moreover, these file systems usually employ a one-size-fits-all replication protocol, which may not be able to provide optimized replication performance for   different write scenarios.

In addition,  there could be heavy accesses to the files by a large number of clients simultaneously.  Most file operations, such as creating, appending, or deleting a file  would require updating the file metadata. Therefore,  a single node  that stores all the file metadata could easily become the performance or storage bottleneck due to the hardware limits ~\cite{hdfs, moosefs}. One can resolve this problem  by  employing a  separate cluster to store the metadata, but most existing works ~\cite{Brandt} on this path would require rebalancing the storage nodes during capacity expansion, which could bring significant degradation on read/write performance.

Lastly, in spite of the fact that having a POSIX-compliant file system interface can greatly simplify the development of the upper level applications, the strongly consistent semantics defined in POSIX I/O standard can also drastically affect the performance.  Most POSIX-compliant file systems  alleviate this issue by providing relaxed  POSIX semantics, but the atomicity requirement between the inode and dentry of the same file can still limit their performance on  metadata operations.

To solve these problems,  in this paper, we propose Chubao File System (CFS), a distributed file system designed for large scale container platforms. CFS is written in Go and the code is available at \textit{\url{https://github.com/ChubaoFS/cfs}}.  Some key features  include:\\

 \noindent  \textit{- General-Purpose and High Performance  Storage Engine}. CFS provides a \textit{general-purpose} storage engine to efficiently store both large and small files with optimized performance on different file access patterns.  It utilizes the punch hole interface in Linux~\cite{punch}  to \textit{asynchronously} free the disk space occupied by the deleted small files, which  greatly simplifies the engineering work of dealing with small file deletions.\\

 \noindent \textit{- Scenario-Aware Replication}.
Different from any existing open source solution that only allows a single replication protocol at any time~\cite{hdfs, ceph, rados},  CFS adopts two strongly consistent replication protocols based on different write scenarios (namely, \textit{append} and \textit{overwrite})  to improve the replication performance.\\

 \noindent \textit{- Utilization-Based Metadata Placement}.
CFS employs a separate cluster to store and distribute the file metadata across different storage nodes based on the memory usage. One advantage of this utilization-based placement is that it does not require any metadata rebalancing during capacity expansion.  Although a similar idea has been used for chunk-server selection in MooseFS~\cite{moosefs}, to the best of knowledge, CFS is the first open source solution to apply this technique for metadata placement. \\

 \noindent \textit{- Relaxed  POSIX Semantics and Metadata Atomicity}. In a POSIX-compliant distributed file system,  the behavior of serving multiple processes on multiple client nodes should be the same as the behavior of a local file system serving multiple processes on a single node with direct attached storage. CFS provides POSIX-compliant APIs. However, the POSIX consistency semantics, as well as the atomicity requirement between the inode and dentry of the same file,  have been  carefully relaxed in order to better align with the needs of applications and to improve the system performance.  

\section{Design and Implementation}
\label{arch}

\begin{figure}[t!]
\begin{center}
\includegraphics[page=1,width=.46\textwidth]{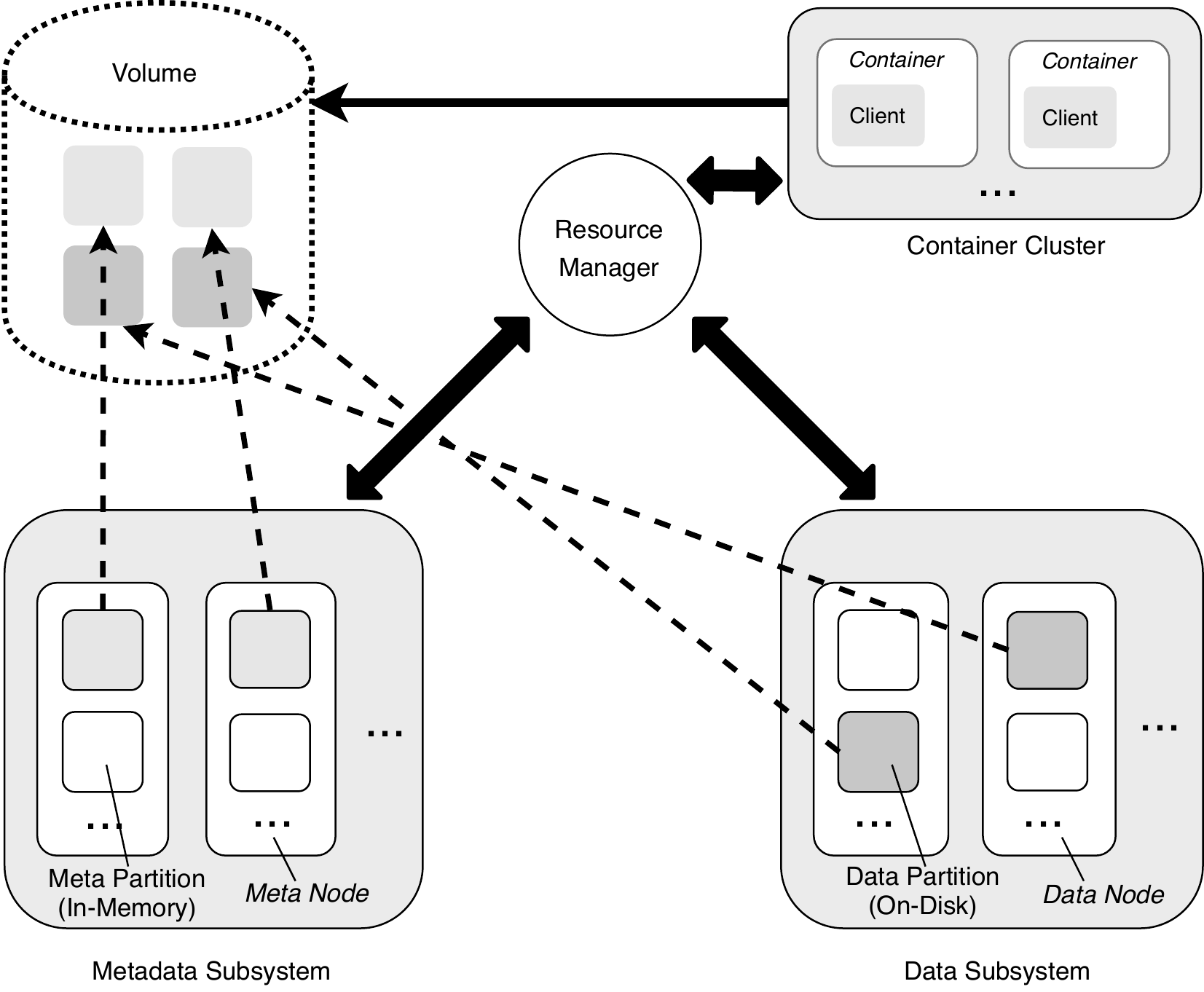}
\caption{Architecture of CFS.}
\label{fig:arch}
\end{center}
\end{figure}

As shown in Figure~\ref{fig:arch}, CFS  consists of a \textit{metadata subsystem}, a \textit{data subsystem}, and a  \textit{resource manager},  and can be accessed by different \textit{clients} as a set of application processes  hosted on the containers. 

The metadata subsystem  stores the file metadata, and consists of a set of \textit{meta nodes}.  Each meta node consists of  a set of \textit{meta partitions}. 
The data subsystem stores the file contents, and  consists of a set of \textit{data nodes}.  Each data node consists of a set of  \textit{data partitions}.  We will give more details about these two subsystems in the following sections.

The volume is a logical concept in CFS and consists of a set of meta partitions and data partitions. Each partition can only be assigned to a single volume. 
From a client's perspective, the volume can be viewed as a file system instance that  contains data accessible by containers.  
A volume can be mounted to multiple containers  so that files can be shared among different clients simultaneously. It needs to be created at the very beginning before the any file operation. 

 The resource manager  manages the file system by processing  different types of tasks  (such as creating and deleting partitions, creating new volumes, and adding/removing nodes). It also keeps track of the status such as the memory and disk utilizations, and liveness of the meta and data nodes in the cluster.  The resource manager has multiple replicas, among which the strong consistency is maintained by a consensus algorithm such as Raft~\cite{raft}, and persisted to a key-value store such as RocksDB\footnote{\url{https://rocksdb.org/}} for backup and recovery.  

\subsection{Metadata Storage}
\label{sec:meta}
The metadata subsystem can be considered as a distributed in-memory datastore of  the file metadata.

\subsubsection{Internal Structure}
The metadata subsystem consists of a set of meta nodes, where each meta node can have hundreds of meta partitions.  

Each meta partition stores the \textit{inodes} and \textit{dentries} of the files from the same volume in memory, and employs two b-trees called \textit{inodeTree}  and \textit{dentryTree}  for fast lookup.  The  \textit{inodeTree} is indexed by the inode id, and the  \textit{dentryTree}  is indexed by the parent inode id and dentry name.   

The  code snippet below shows the definitions of the meta partition, the inode and the dentry in CFS. 


{\small
\begin{lstlisting}[language=C]
type metaPartition struct {
	config        *MetaPartitionConfig
	size          uint64 
	dentryTree    *BTree     // btree for dentries
	inodeTree     *BTree.    // btree for inodes
	raftPartition   raftstore.Partition
	freeList      *freeList  // free inode list
	vol			  *Vol
	...                      // other fields
}

type inode struct {
	inode        uint64 // inode id
	type         uint32 // inode type
	linkTarget   []byte // symLink target name
	nLink        uint32 // number of links 
	flag         uint32 
	...                  // other fields 
}

type dentry struct {
	parentId uint64 // parent inode id
	name      string // name of the dentry
	inode     uint64 // current inode id
	type      uint32 // dentry type
}

\end{lstlisting}
}
 \subsubsection{Raft-based Replication}
The replication during file write is performed in terms of meta partitions.
The strong consistency among the replicas is ensured by a  revision of the  Raft consensus protocol~\cite{raft} called the  MultiRaft\footnote{\url{https://github.com/cockroachdb/cockroach/tree/v0.1-alpha/multiraft}}, which has the advantage of reduced  heartbeat  traffic on the network comparing to the original version.  

\subsubsection{Failure Recovery}
The  meta partitions cached in the memory are  persisted  to the local disk by snapshots and  logs ~\cite{raft} for backup and recovery. Some techniques such as log compaction are used to reduce the log files sizes and shorten the recovery time.

It is worth noting that, a  failure  that happens during a metadata operation could result an \textit{orphan inode}  with which has no dentry to be associated. The memory and disk space occupied by this inode can be hard to free.  To minimize the chance of this case to happen, the client always issues a retry after a failure until the request succeeds or the maximum retry limit is reached. 

\subsection{Data Storage}
\label{sec:data}
The data subsystem is optimized for the storage of both large and small files, which can be accessed in a sequential or random fashion.  

 
 
 

\subsubsection{Internal Structure}
The data subsystem consists of a set of data nodes, where each data node has a set of data partitions.

\begin{figure}[t!]
\begin{center}
\includegraphics[page=1,width=.44\textwidth]{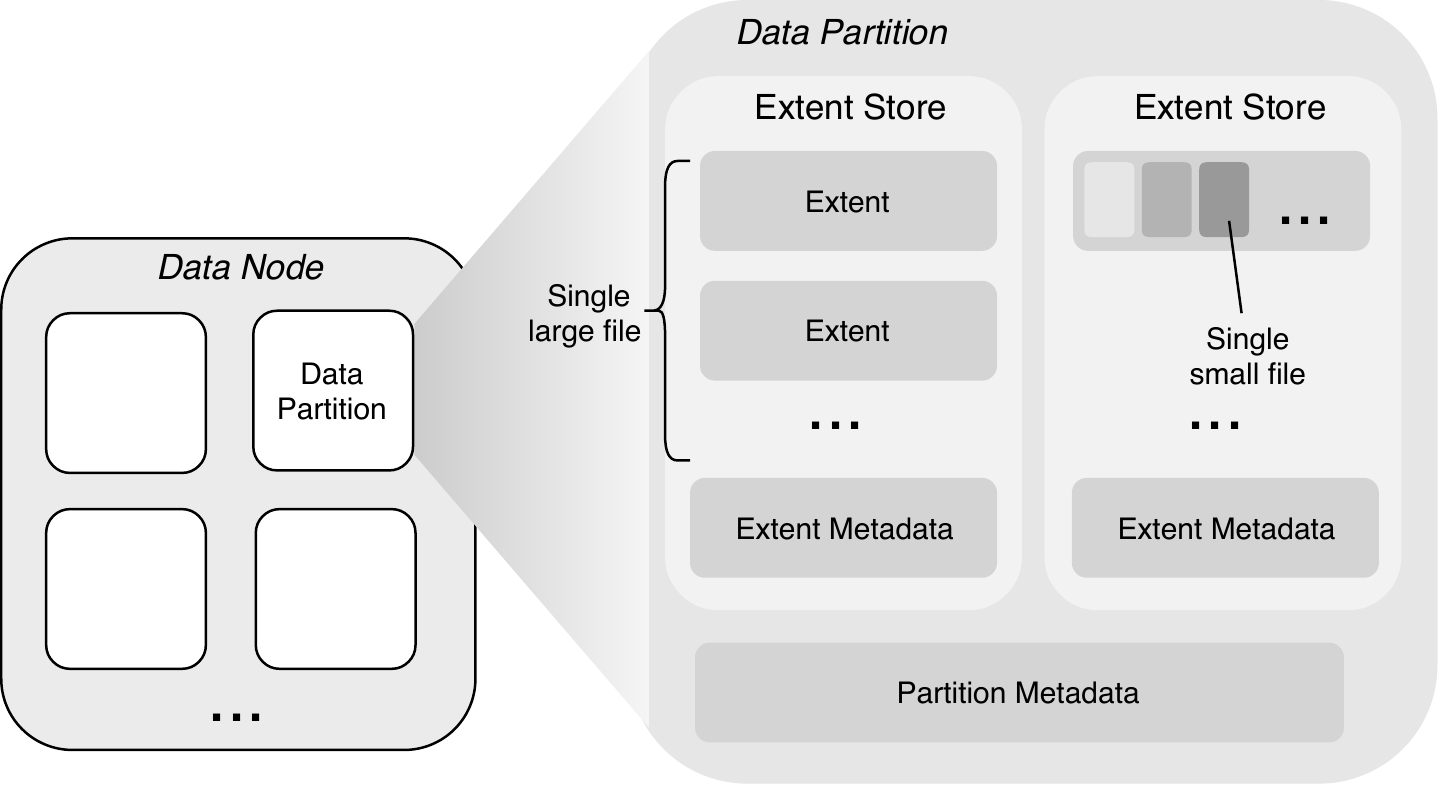}
\caption{Internal structure of a data partition.}
\label{fig:data}
\end{center}
\end{figure}

  Each data partition stores a set of partition metadata such as partition id and the replica addresses. It also has  a  storage engine called the \textit{extent store} (see Figure~\ref{fig:data}), which is constructed by a set of storage units called   \textit{extents}. The CRC of each extent is cached in the memory to speed up the check for data integrity.   The code snippet below shows the structure of the data partition in CFS. 

{\small
\begin{lstlisting}[language=C]
type dataPartition struct {
	clusterID       string
	volumeID        string 
	partitionID     uint64 
	partitionStatus int
	partitionSize   int
	replicas        []string // replica addresses
	disk            *Disk
	isLeader        bool
	isRaftLeader    bool
	path            string
	extentStore     *storage.ExtentStore
	raftPartition   raftstore.Partition
	...                      // other fields
}
\end{lstlisting}
}

Large files and small files are stored in different ways, and a threshold $t$ (128 KB by default) is used to determine if the file should be considered as "small file" or not, i.e., the file whose size is less than or equal to $t$ will be treated as "small file". The value of $t$ is configurable at the startup and usually aligned with the packet size during the data transfer to avoid the packet assembly or splitting. 

\subsubsection{Large File Storage}
For large files, the contents are stored as a sequence of one or multiple extents, which can be  distributed across different data partitions on different data nodes.  
Writing a new file to the extent store always causes the data to be written at the zero-offset of a new extent, which eliminates the need for the offset within the extent. The last extent of a file  does not need to fill up its size limit by padding (i.e., the extent  does not have holes), and never stores the data from other files.

\subsubsection{Small File Storage and Punch Holes}
The contents of multiple small files are aggregated and stored in a single extent, and the physical offset of each file content in the extent is recorded in the corresponding meta node.  CFS relies on the punch hole interface, \textit{fallocate()}\footnote{\url{http://man7.org/linux/man-pages/man2/fallocate.2.html}},  to \textit{asynchronous} free the disk space occupied by the to-be-deleted file. The advantage of this design is to eliminate the need of implementing a garbage collection mechanism and therefore avoid to employ a mapping from logical offset to physical offset  in an extent~\cite{haystack}.  Note that this is different from deleting large files, where  the extents of the file can be removed directly from the disk.

\subsubsection{Scenario-Aware Replication}
\label{sec:rep}
Although one can simply apply a Raft-based replication to ensure the strong consistency, CFS employs two replication protocols for the data subsystem to achieve  a better tradeoff on the performance and code reusability.  

The replication is performed in terms  of partitions during file writes. 
Depending on the file write pattern, CFS adopts different strongly consistent replication strategies. Specifically, for sequential write (i.e., \textit{append}), we use the primary-backup replication~\cite{chain, rados},  and for overwrite, we employ a MultiRaft-based replication protocol similar to the one used  in the metadata subsystem.

One reason is that  the  primary-backup replication is not suitable for overwrite as the replication performance may need to be compromised. Specifically, during overwrite, at least one new extent will be created to store the new file content(s), and some of the original extents will be \textit{logically} split into several fragmentations, which are usually linked together like a linked list. In this linked list, the pointer to the original fragmentations will be replaced by the ones associated with the newly created extent(s). As more and more file contents being overwritten, eventually there will be too many  fragmentations  on the data partitions that requires \textit{defragmentation}, which could significantly affect the replication performance.

The other reason is that the MultiRaft-based replication is known to have the write amplification  issue as it introduces extra IO of  writing the log files, which could directly affect the read-after-write performance. However,  because our applications and microservices usually have much less overwrites than sequential writes,   this performance issue  can be tolerable. 



\subsubsection{Failure Recovery}
Due to the existence of two different replication protocols, when a failure on a replica is discovered, we first start the recovery process  in the primary-backup replication by checking and aligning all extents. Once this processed is finished, we then start the  recovery process  in our MultiRaft-based replication.

It should be noted that, during a sequential write, the stale data is allowed on the partitions, as long as it is never returned to the client. This is because that, in this case, the commitment of  the data at an offset also indicates the commitment of all the data before that offset. As a result, we can use the offset to indicate the portion of the data that has been committed by all replicas, and the leader always returns the largest offset that has been committed by all the replicas to the client, who will then update this offset at the corresponding meta node. When the client requests a read, the  meta node  only provides  the address of a replica with the offset of the data that has been committed by all replicas, regardless the existence of the stale data that has not  been fully committed. 

Therefore, there is no need to recover the inconsistent portion of the data on the replicas. If a client sends a write request for a $k$ MB file, and only the first $p$ MB have been committed by all replicas, then the client will resend a write request for the remaining $k - p$ MB data to the extents in different data partitions/nodes. This design  greatly simplifies our implementation of maintaining the strong replication consistency when a file is sequentially written to CFS.


\subsection{Resource Manager} 
 The resource manager  manages the file system by  processing  different types of tasks.  
\subsubsection{Utilization-Based  Placement}
One important task is to distribute  the file metadata and contents  among different nodes.  In CFS, we adopt a \textit{utilization-based placement}, where the file metadata and contents are  distributed across the storage nodes based on the memory/disk usage. This  greatly simplifies the resource allocation problem in a distributed file system.  

To the best of our knowledge, CFS is the first open source solution to employ this idea for metadata placement.  
Some commonly-used metadata placements, such as \textit{hashing} and \textit{subtree partition}~\cite{Brandt} usually require disproportionate amount of metadata to be moved when adding servers. Such data migration could be a headache in a container environment as capacity expansion usually needs to be finished in a short period of time. Other approaches such as \textit{lazyhybrid}~\cite{Brandt} and \textit{dynamic subtree partition}~\cite{ceph2} move the data lazily or employ a proxy to hide the data migration latency.  But these solutions also bring extra amount of engineering work for future development and maintenance. Different from these approaches,  the utilization-based  placement, despite its simplicity,  does not require any rebalancing  of the data when adding new storage nodes during capacity expansion. In addition, because of the uniformed distribution,  the chance of  multiple clients accessing the data on the same storage node simultaneously can be reduced, which could potentially improve the performance stability of the file system.
Specifically, our utilization-based placement  works as follows:

First, after creating a volume, the client  asks the resource manager for a certain number of available meta and data partitions. These partitions are usually the ones  on the nodes with the lowest memory/disk utilizations. However, it should be noted that, when writing a file, the client  simply selects the meta and data partitions  in a random fashion  from the ones allocated by the resource manager. The reason that the client does not adopt similar utilization-based approach is to avoid the communication with the resource manager in order to obtain the update-to-date  utilization information of each allocated node. 

Second, when the resource manager  finds that all the partitions in a volume is about to be full,  it  automatically adds a set of new  partitions to this volume.  These partitions are usually the ones on the nodes with the lowest memory/disk utilizations.  Note that, when a  partition is full, or a threshold (i.e.,  the number of files on a meta partition or the number of extents on a data partition) is reached, no new data can be stored on this partition, although it can still be modified or deleted. 

\subsubsection {Meta Partition Splitting}

{\small
\begin{algorithm}[!t]
\caption{Splitting Meta Partition}\label{split}
\begin{algorithmic}[1]
\Procedure{Partitioning}{}
\State $\textit{mp} \gets \text{current meta partition}$
\State $\textit{c} \gets \text{current cluster}$
\State $\textit{v} \gets \text{cluster.getVolume(mp.volName);}$ 
\State $\textit{maxPartitionID} \gets \text{v}.\text{getMaxPartitionID()}$
\If {$\textit{metaPartition.ID} < \textit{maxPartitionID}$}
\Return
\EndIf
\If {$\textit{mp.end == math.MaxUint64}$}
		\State $\textit{end $\gets$ maxInodeID + $\Delta$}$ \Comment $\text{curoff the inode range}$
		\State $\textit{mp.end $\gets$ end}$
		\State $\textit{task} \gets \text{newSplitTask(\textit{c.Name, mp.partitionID, end})}$
		\State $\text{}$
		\State $\text{c.addTask(\textit{task})}$ \Comment $\text{sync with the meta node}$
		\State $\text{}$

		\State $\text{\textit{c}.updateMetaPartition(\textit{mp.volName, mp})}$ 
		\State $\text{\textit{c}.createMetaPartition(\textit{mp.volName, mp.end+1})}$
\EndIf
\EndProcedure
\end{algorithmic}
\end{algorithm}
}

The resource manager also needs to handle a special requirement when splitting a meta partition.
 In particular,  if  a meta partition is about to reach its  upper  limit of the number of stored inodes and  dentries,  a splitting task needs to be performed to ensure that the inode ids stored at the newly created partition are unique from the ones stored at the original partition.
  
  The  pseudocode of our solution is given in Algorithm~\ref{split}, in which,  the resource manager first  cuts off the inode range of the meta partition in advance at a upper bound $end$, a value greater than largest inode id used so far (denoted as $maxInodeID$), and then sends a  \textit{split} request to the meta node  to (1) update the inode id range from $1$ to  $end$  for the original meta partition, and (2) create a new meta partition with the inode range from $end + 1$ to  $\infty$. 
 As a result, the inode range for these two meta partitions becomes $[1, end]$ and $[end + 1, \infty]$, respectively. If another file needs to be created, then its inode id will be chosen as $maxInodeID + 1$ in the original meta partition, or $end + 1$ in the newly created meta partition. 
 The $maxInodeID$ of each meta partition can be obtained by the periodical communication between the resource manager and the meta nodes.

\subsubsection{Exception Handling}
When a request to a meta/data partition times out (e.g., due to  network outage), the remaining replicas are marked as read-only. 
When a meta/data partition is no longer available (e.g., due to hardware failures), all the data on this partition will eventually be migrated to a new  partition manually. This unavailability is identified by the multiple failures reported by the  node.

\subsection{Client} 
\label{sec:client}


 

The client has been integrated with FUSE\footnote{\url{https://github.com/libfuse/libfuse}} to provide a file system interface in the user space. 
The client process runs entirely in the user space with its own cache. 
 
To reduce the communication with the resource manager,  the client caches the addresses of the available meta and data partitions assigned to the mounted volume, which can be  obtained at the startup, and periodically synchronizes this available partitions with the resource manager. 

To reduce the communication with the meta nodes, the client also  caches the returned inodes and dentries  when creating new files, as well as the data partition id, the extent id and the offset, after the file has been written to the data node successfully.  When a file is opened for read/write, the client will force  the cache metadata to be synchronous with the meta node. 
 
To reduce the communication with the data nodes,  the client caches the most recently identified leader. Our observation is that, when reading a file, the client may not know which data node is the current leader because the leader could change after a failure recovery. As a result, the client may try to send the read request to each replica one by one until a leader is identified.  However, since  the leader does not change  frequently,   by caching the last identified leader, the client can have minimized  number of retries in most cases.

 \subsection{Optimizations}
 
 There are mainly two optimizations being applied for the CFS clusters deployed at JD, as explained below:
 
 \subsubsection{Minimizing Heartbeats}
Because our production environment could have a huge number of partitions spread across different meta and data nodes, even with the MultiRaft-based protocol, each node can still get an  explosion of the heartbeats  from the rest in the same Raft group, causing significant communication overheads. To alleviate this, we employ an extra layer of abstraction on the  nodes called \textit{Raft set} to further minimize the number of  heartbeats to be exchanged among the Raft groups. Specifically, we divided all the   nodes  into several Raft set, each of which maintains its own Raft group. When creating a new partition, we prefer to select the replicas from the same Raft set. In this way, each node only needs to exchange  heartbeats with  the nodes in the same Raft set.

\subsubsection{Non-Persistent Connections}
There can be tens of thousands of clients accessing the same CFS cluster, which could cause the resource manager to be overloaded if all their connections are kept alive. To prevent this from happening,  we use  \textit{non-persistent connection} between each client and the resource manager. 

\subsection{Metadata Operations}

\begin{figure*}[t!]
\begin{center}
\includegraphics[page=1,width=.96\textwidth]{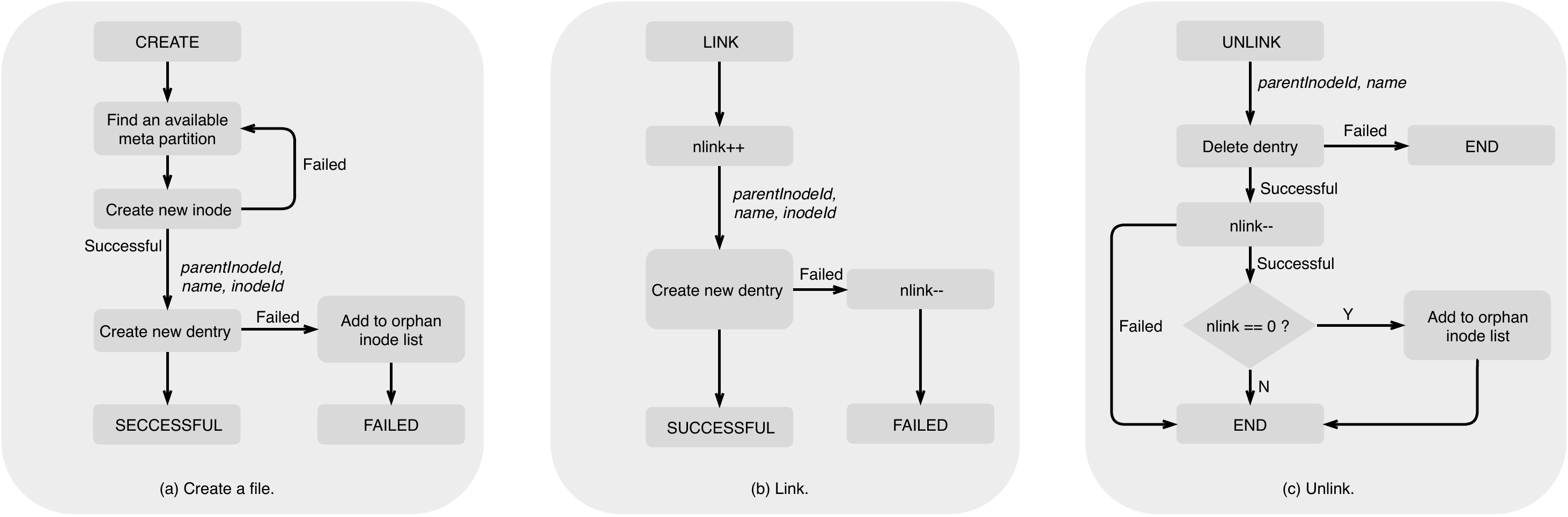}
\caption{Workflows of three commonly used metadata operations.}
\label{fig:meta-flow}
\end{center}
\end{figure*}

%
In most modern POSIX-compliant and  distributed file systems~\cite{ceph}, the  inode and dentry of a file usually reside on the same storage node in order to preserve the directory locality\footnote{Directory locality is a term referring to the phenomenon that the  metadata of files under the same directory are likely to be accessed repeatedly, i.e., when a file metadata is accesses, it is likely that the metadata of the files under the same directory, or the directory metadata will also be accessed.}. 
However, because of the utilization-based metadata placement, in CFS,  the inode and dentry of the same file may be distributed across different metadata nodes. 
As a result,  operating  a file's inode and dentry usually needs to run as distributed transactions, which brings the overheads that could significantly affect the system performance. 

Our tradeoff is to relax this atomicity requirement \textit{as long as  a dentry is always associated with at least one inode}. All the metadata operations in CFS are based on this design principle. The downside is that there is a chance to create \textit{orphan inodes}\footnote{An orphan inode is an inode that has no dentry to be associated with.}, which may be difficult to be released from the memory.  To mitigate this issue,  each metadata-operation workflow in CFS has been carefully designed  to \textit{minimize the chance of an orphan inode to appear}. In practice, a meta node rarely has too many orphan inodes in the memory. But if this happens,  tools like \textit{fsck} can be used to repair the files by the administrator.  

 Figure~\ref{fig:meta-flow} illustrates the workflows of  three  common  metadata operations, which can be explained as follows.
 
 \subsubsection{Create}
 When creating a file,  the client first asks an available meta node to create an inode.  The meta node  picks up the smallest  inode id that has not been used so far in this partition for the newly created inode, and updates  its largest  inode id   accordingly.  Only when  the inode has been successfully created,  the client can ask for creating a corresponding dentry. If a failure happens,  the client will send an \textit{unlink} request, and put the newly created inode  into a local list of orphan inodes, who will  be deleted when the meta node receives an evict request from the client. Note that the inode and dentry of the same file do not need to be stored on the same meta node. 

 \subsubsection{Link}
When linking a file, the client first asks the meta node of the inode  to increase the value of \textit{nlink} (the number of associated links)  by one, and then asks the meta node of the target parent inode  to create a dentry on the same meta partition. If a failure happens when creating the  dentry,  the value of \textit{nlink} will be decreased by one.

 \subsubsection{Unlink}
When unlinking a file, the client first asks  the corresponding meta node to delete  the dentry. Only when this operation succeeds, the client  sends an \textit{unlink} request to the meta node to decrease the value of \textit{nlink} in the target inode by one. When it reaches certain threshold (0 for file and 2 for directory),  the client  puts this inode into a local list of orphan inodes, who will  be deleted when the meta node receives an evict request from the client.  Note that, if decreasing the value of \textit{nlink} fails, the client will perform several retries. If all the retries failed,  this inode will eventually become an orphan inode, and  the administrator may need to manually resolve the issue.

\subsection{File Operations}
CFS has relaxed POSIX consistency semantics, i.e., instead of providing strong consistency guarantees,  it only ensures  sequential consistency for file/directory operations, and does not have any leasing mechanism to prevent multiple clients writing to the same file/directory. It depends on the upper-level application to maintain a more restrict  consistency level if necessary. 

 \subsubsection{Sequential Write}
\begin{figure}[t!]
\begin{center}
\includegraphics[page=1,width=.48\textwidth]{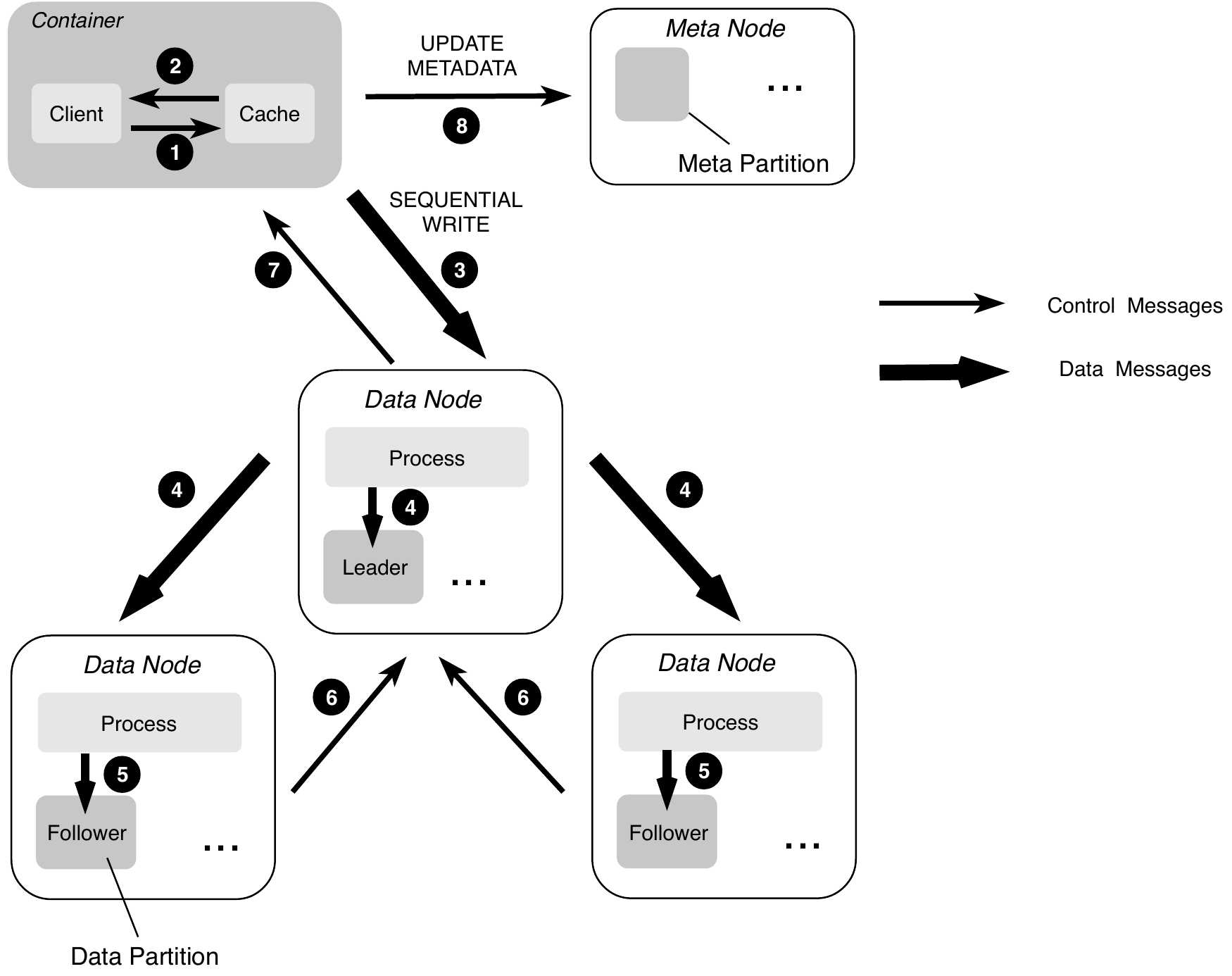}
\caption{ Workflow of sequential write. }
\label{fig:app}
\end{center}
\end{figure}

As shown in Figure~\ref{fig:app}, to sequentially write a file, the client first randomly chooses the available data partitions from the cache,  then continuously sends a number of fixed sized  packets (e.g., 128 KB) to the leader, each of which includes the addresses of the replicas, the target extent id, the offset in the extent, and the file content.  The addresses of the replicas are provided as an array by the resource manager and cached on the client side. The  indices of the items in this array tell the order of the  replication, i.e., the replica whose address at the  index zero is the leader. Therefore, the client can always send a write request to the leader without introducing extra communication overhead, and  as mentioned in the previous section, a primary-backup  replication will be performed by following this order.  Once the client receives the commit from the leader, it updates the local cache immediately, and synchronizes with meta node periodically or upon receiving a system call, \textit{fsync()}\footnote{\url{http://man7.org/linux/man-pages/man2/fdatasync.2.html}}, from the upper level application.

\subsubsection{Random Write}
\begin{figure}[t!]
\begin{center}
\includegraphics[page=1,width=.42\textwidth]{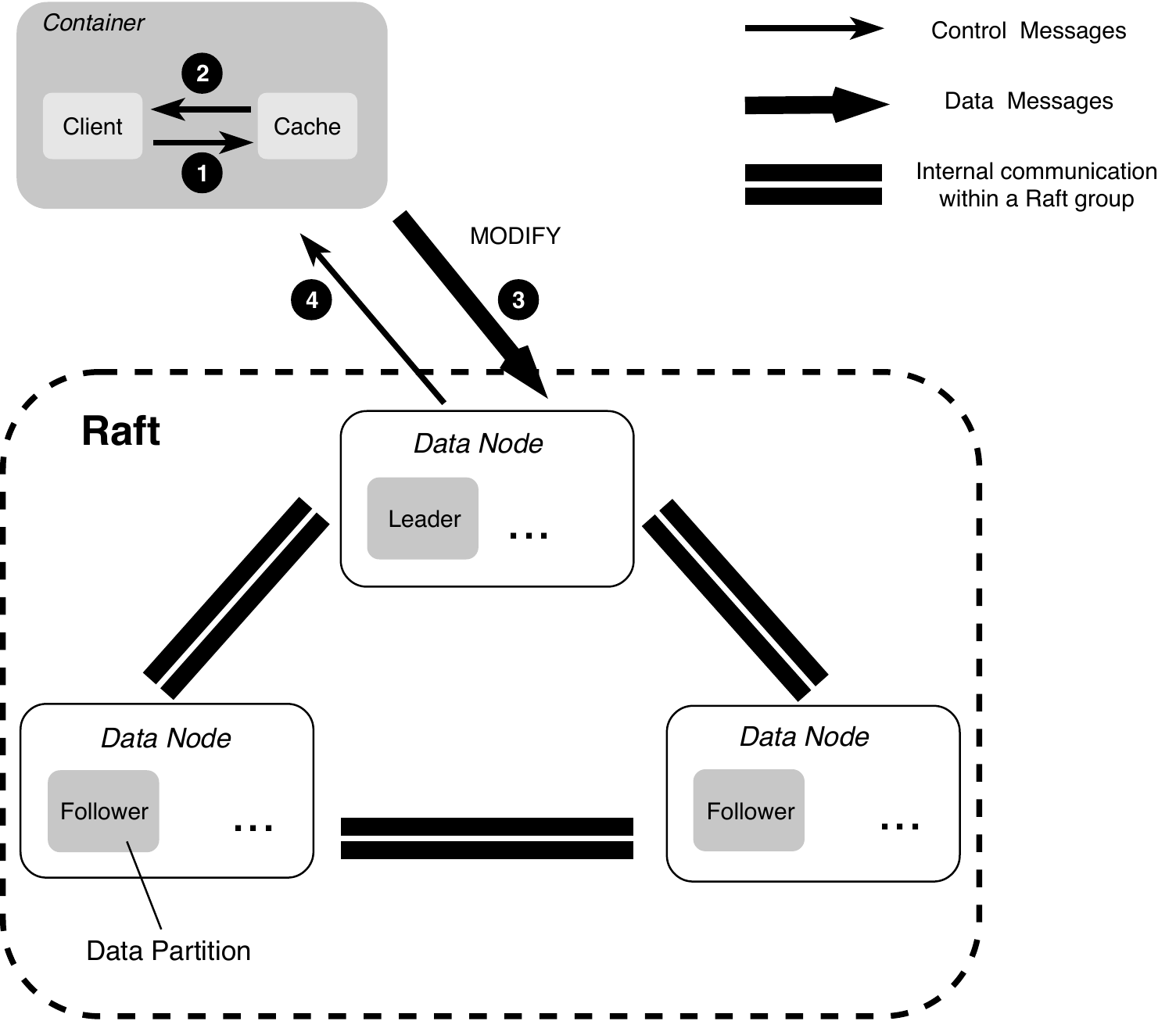}
\caption{ Workflow of overwriting an existing file (no appending).  }
\label{fig:modify}
\end{center}
\end{figure}

The random write in CFS is \textit{in-place}.
To randomly write a file, the client  first  uses the offsets of the original data and the new data to  figure out the portions of data to be appended and the portions of data to be overwritten, and then processes them separately. In the former case, the client  sequentially writes the file as described previously. In the latter case,  as shown in Figure~\ref{fig:modify},  the offset of the file on the data partition does not change. 

\subsubsection{Delete}
 The delete operation is asynchronous. To delete a file, the client sends a delete request to the corresponding meta node. The meta node, once receives this request, updates the value of  \textit{nlink}  in  the target inode. If this value reaches certain threshold (0 for file and 2 for directory),  the  target inode will be marked as deleted (see the inode structure given in Section~\ref{sec:meta}). Later on, there will be a separate process to clear up this inode and communicate with the data node to delete the file content.
 
\subsubsection{Read}
A read can only happen at the Raft leader (note that the primary-backup group leader and raft group leader could be different). To read a file, the client  sends  a read request to the corresponding data node. This request is constructed by the data from the client cache, such as the data partition id, the extent id,  the offset of the extent, etc.



\section{Discussion of Design Choices}
\label{sec:design}
In this section, we highlight some of the design choices we have made when building CFS.
\subsection{Centralization vs. Decentralization}
Centralization and decentralization are two design paradigms for distributed systems ~\cite{ali}. Although the former one~\cite{gfs, hdfs}  is relatively easy to implement, the single master could become a bottleneck in the sense of  the scalability. In contrast, the latter~\cite{dynamo} one is generally more reliable but also more complex to implement.


 In designing CFS, we choose centralization over decentralization mainly for the reason of its simplicity. 
However,  a single resource manager that stores all  the  file metadata  limits the scalability  of the  metadata operations, which could make up as much as half of typical file system workloads ~\cite{compare}. For this reason, we employ a separate cluster to store the metadata, which drastically improves the scalability of the entire file system. From this perspective, CFS is designed in a way to minimize the involvements of the resource manager so that it has less chance to become a bottleneck. Admittedly, even with these efforts, the  scalability of the resource manager  could still be limited by its memory and disk space. But based on our  experience, this never becomes an issue.  

\subsection{Separate Meta Node vs. Metadata on Data Node}
In some distributed file systems, the file metadata and contents are stored on the same machine~\cite{ceph, Morris}. There are also some  distributed file systems where the metadata is managed separately by specialized metadata servers~\cite{hdfs, Gibson}.

In CFS, we choose to have a separate meta node and maintain all the file metadata in the memory for fast access. As a result, we can select memory-intensive machines for the meta nodes, and disk-intensive machines for the data nodes for  cost-effectiveness. Another advantage of this design is the flexibility of deployment. In case  a machine has both large memory and disk spaces, we can always deploy the meta node and data node physically together.

\subsection{Consistency Model and Guarantees}
In CFS,  the storage layer and file system layer have different consistency models. 

The storage engine guarantees the strong consistency among the replicas through either primary-backup or Raft-based replication protocols.  This design decision is based on the observations that the former one  is not suitable for overwrite as the replication performance needs to be compromised, and the latter one has write amplification issue as it introduces extra IO of writing the log files.   

The file system itself, although provides POSIX-compliant APIs, has selectively relaxed POSIX  consistency semantics in order to better align with the needs of applications and to improve system performance. 
For example, the semantics of POSIX  defines  that writes must be strongly consistent, i.e, a write is required to block application execution until the system can guarantee that any other read will see the data that was just written.  While this can be easily accomplished locally, ensuring such strong consistency on a distributed file system is very challenging due to the degraded performance associated with lock contentions/synchronizations. CFS  relaxes the POSIX consistency in a way that provides consistency when different clients modify non-overlapping parts of a file, but it does not provide any consistency guarantee if two clients try to modify the same portion of the  file. 
This design is based on the fact that in a containerized environment, there are many cases where the rigidity of POSIX semantics is not strictly necessary, i.e.,  
the applications seldom rely on the file system to deliver full strong consistency, and two independent jobs rarely write to a common shared file on a multi-tenant system. 

\section{Evaluation}
\label{exp} 

Ceph is a distributed file system that has been widely used on   container platforms.
In this section, we perform a set of experiments to compare CFS with Ceph from various aspects.  

\subsection{Experiment Setup}

\begin{table}
\small
\center
\caption{System specification. }
\begin{tabular}{l | l}
\toprule
\textbf{Processor Number}  & Xeon E5-2683V4\\ 
\textbf{Number of Cores} & 16   \\
\textbf{Max Turbo Frequency} & 3.00 GHz \\
\textbf{Processor Base Frequency} & 2.10 GHz  \\ 
\textbf{Network Bandwidth} & 1000Mbps  \\ 
\textbf{Memory} & DDR4 2400MHZ, 8 $\times$ 32 GB \\
\textbf{Disk} & 16 $\times$ 960 GB SSD  \\ 
\textbf{Operating System} & Linux 4.17.12  \\  \bottomrule
\end{tabular}
\label{exp:configure}  
\end{table}

The  specifications of the machines used in our experiments are given in Table ~\ref{exp:configure}. 
For CFS, we deploy the meta nodes and data nodes on the same cluster of 10 machines, and a single resource manager with 3 replicas.  Each machine has 10 meta partitions and 1500 data partitions.
For Ceph, we have similar setup, where  the object
storage devices (OSD) and metadata server (MDS) are deployed on the same cluster of 10 machines. Each machine has 16 OSD processes and 1 MSD process. The Ceph version is 12.2.11, and the storage engine is configured as the \textit{bluestore} with the TCP/IP network stack.  Unless otherwise stated, both CFS and Ceph use default configurations in the following experiments. 

\subsection{Metadata Operations}
\begin{table}
\small
\center
\caption{Description of the tests in \textit{mdtest}.}
\begin{tabular}{l | l}
\toprule
\textbf{DirCreation}  & Create a directory\\ 
\textbf{DirStat} & List all the files in the current directory    \\
\textbf{DirRemoval} & Remove a directory\\
\textbf{FileCreation} & Create a file \\ 
\textbf{FileRemoval} & Remove the file attributes  \\ 
\textbf{TreeCreation} & Create a directory with multiple files as a \\
& tree structure \\
\textbf{TreeRemoval} & Remove a directory with multiple files as a\\ 
&  tree structure \\  \bottomrule
\end{tabular}
\label{exp:meta}  
\end{table}

\begin{figure*}[t!]
\begin{center}
\includegraphics[page=1,width=.8\textwidth]{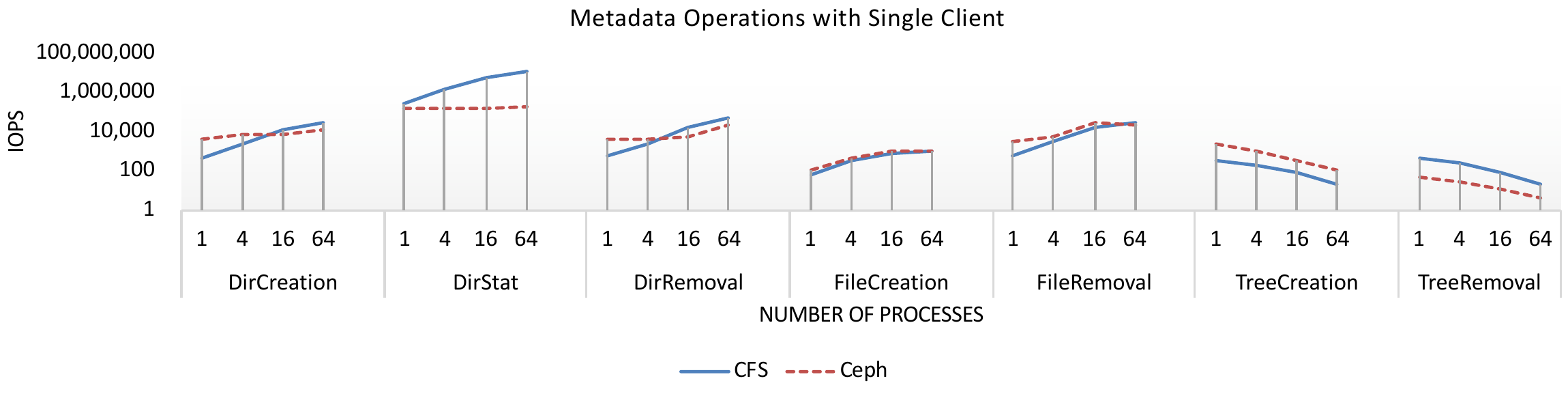} 
\caption{IOPS when a single client operates the file metadata.}
\label{fig:meta1}
\end{center}
\end{figure*}

\begin{figure*}[t!]
\begin{center}
\includegraphics[page=1,width=.8\textwidth]{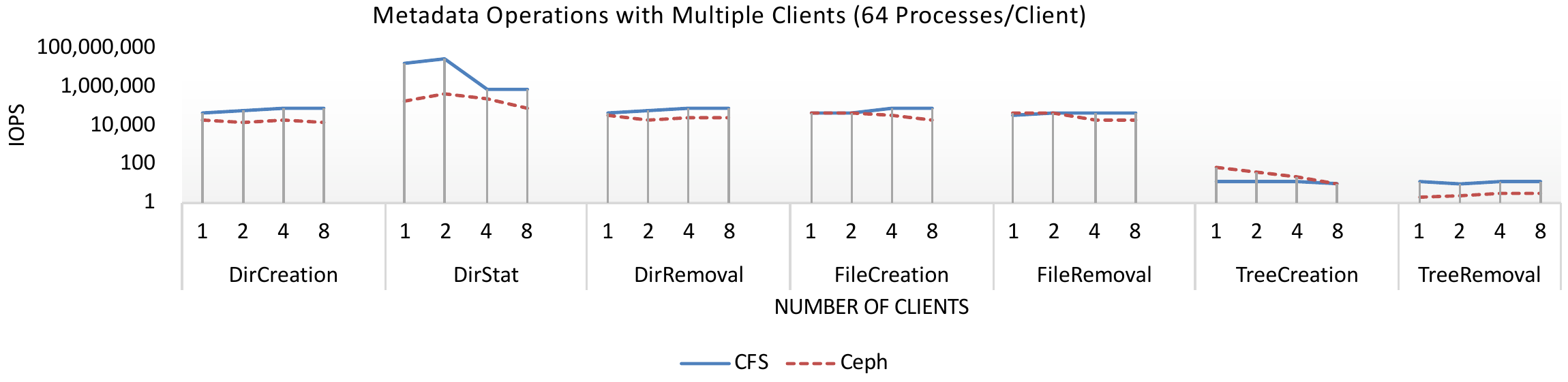} 
\caption{IOPS when multiple clients operate  the file metadata.}
\label{fig:meta2}
\end{center}
\end{figure*}

\begin{table}
\small
\center
\caption{IOPS for the metadata operations with 8 clients where each client has 64 processes.}
\begin{tabular}{|l | r | r | r | r |} 
\toprule
\textbf{Test Name}  & CFS (multi) & Ceph (multi) & \% of Improv.\\ \midrule
\textbf{DirCreation}  & 83,729 & 16,627 & 404\\ 
\textbf{DirStat} & 875,867 & 91,050 &   862 \\
\textbf{DirRemoval} & 94,235 & 23,807 & 296\\
\textbf{FileCreation} & 85,556 & 21,919 & 290  \\ 
\textbf{FileRemoval} & 50,119 & 22,573 &  122\\ 
\textbf{TreeCreation} & 10 & 11 & -9 \\
\textbf{TreeRemoval} & 12 & 3 &300 \\  \bottomrule
\end{tabular}
\label{exp:meta2}  
\end{table}

In  evaluating the performance and scalability of the metadata subsystem, we focus on the tests of 7 commonly used metadata operations from  \textit{mdtest}\footnote{\url{https://github.com/hpc/ior}}.  The description is given in Table~\ref{exp:meta}.  Note that the \textit{TreeCreation} and \textit{TreeRemoval} tests mainly focus on operating directories  as  non-leaf nodes in the tree structure. 


Figure~\ref{fig:meta1} plots the IOPS of these tests in a single-client environment with different number of processes, and  Figure~\ref{fig:meta2} plots the IOPS of the same set of tests in a multi-client environment, where each client runs 64 processes. Note that the Y-axis uses the \textit{logarithmic scale} for better illustration with different test results. 

It can be seen that,  when there is only a single client with a single process, Ceph outperforms CFS in 5 out of 7 tests (except the \textit{DirStat}  and \textit{TreeRemoval} tests), but as the number of clients and processes increase, CFS starts to catch up.  When it reaches to 8 clients (each running 64 processes), CFS outperforms Ceph in 6 out of 7 tests (except the \textit{TreeCreation} test). The detailed IOPS  of the tests with 8 clients is given in Table~\ref{exp:meta2}, which shows about 3 times performance boost  on the average. From this result we can see that, as the number of clients and processes gets increased, the performance benefits from the utilization-based metadata placement in CFS can probably outweigh the advantage from the directory locality-aware metadata placement  in Ceph.

There are  a few  observations from the results of \textit{DirStat},   \textit{TreeCreation} and \textit{TreeRemoval}, as explained below.

In the \textit{DirStat} test, CFS exhibits better performance than Ceph in both  cases (i.e., single-client and multi-client). This is mainly because that they handle the \textit{readdir}  request in different ways.   In Ceph, each \textit{readdir} request is  followed by a set of \textit{inodeGet} requests to  fetch all the inodes in the current directory from different MDS. The requested inodes are usually cache in the MDS for the fast access in the future. However,  in CFS,  these \textit{inodeGet} requests are replaced  by a \textit{batchInodeGet} request in order to reduce the communication overheads, and the results  are cached  at the client side  so that  the successive  requests can be quickly responded without further communications with the same set of  meta nodes. The sudden performance drops in the multi-client case (see Figure~\ref{fig:meta2}) can be caused by the client cache misses in CFS.   

In the \textit{TreeCreation} test,  Ceph performs better than CFS in the single-client case. But as more clients get involved, the performance gap between them gets closer. This can be caused by the fact that, when there are only  a few clients,  the benefits given by the directory locality such as reusing the cached dentries and inodes on the same MDS gives Ceph a better performance.  However, as more clients get involved, the increasing pressure on certain MDS requires Ceph  to dynamically place the metadata of  files  under the same directory into different MDSs and  redirect  the corresponding requests to the proxy MDSs~\cite{ceph2}, which incurs extra  overheads and closes the  gap between Ceph and CFS.

In the \textit{TreeRemoval} test, CFS  gives better performance than Ceph in both cases. Similar to the \textit{DirStat} test,  the way CFS handles the \textit{readdir} request can be one of the reasons for such results.   In addition,  when there are only a few  clients, 
the requests of deleting the file metadata may need to be queued in Ceph as the metadata of the files under the same directory are usually stored on a single MDS; and as more clients get involved, the potential benefits given by the directory locality in Ceph may be reduced since the file metadata may have been distributed across  different MDSs because of the rebalancing triggered in the previous tests.

\subsection{Large Files}

\begin{figure}[t!]
\begin{center}
\includegraphics[page=1,width=.48\textwidth]{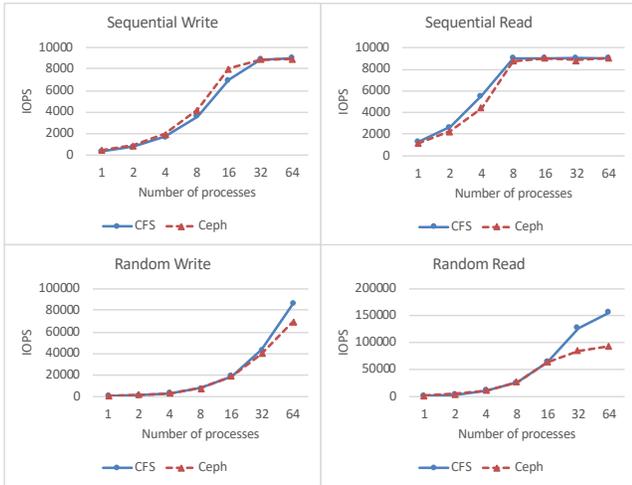}
\caption{ IOPS when different number of processes, in a single client, operate files with different access patterns. Eeach process operates a separate 40 GB file.}
\label{fig:single}
\end{center}
\end{figure}

For large file operations, we first  look at  the results with different number of processes in a single-client environment, where each process operates a separate 40 GB file. We use \textit{fio}\footnote{\url{https://github.com/axboe/fio}} (with \textit{direct IO} mode) to generate various types of workloads. In both setups of CFS and Ceph,  the  clients and servers are deployed on different machines.  In addition, two parameters  need to be tuned in Ceph in order to obtain the optimal performance, namely, the \textit{osd$\_$op$\_$num$\_$shards} and \textit{osd$\_$op$\_$num$\_$threads$\_$per$\_$shard}, which  control the number of queues and the number of threads to process the queues. We set them to $6$ and $4$   respectively. Increasing any of these values  further  will cause degraded write performance due to the high  CPU pressure.
 
 As shown in Figure~\ref{fig:single},  the performance under different processes are quite similar, with the exception that in the random read/write tests, CFS has higher IOPS when the number of processes is greater than 16. 
 
 This is probably  due to the following reasons.
 First, each MDS of Ceph only caches a portion of the file metadata in its memory.  In the case of random read,  the cache miss rate can be increased dramatically as the number of processes increases, causing frequent disk IOs. In contrast, each meta node of CFS caches all the file metadata in the memory to avoid the expensive disk IOs. Second, the overwrite in CFS is in-place, which does not need to update the file metadata. In contrast, the overwrite in Ceph usually needs to walk through multiple queues, and only after the data  and metadata have been persisted and synchronized, the commit message can be returned to the client.  

\begin{figure}[t!]
\begin{center}
\includegraphics[page=1,width=.48\textwidth]{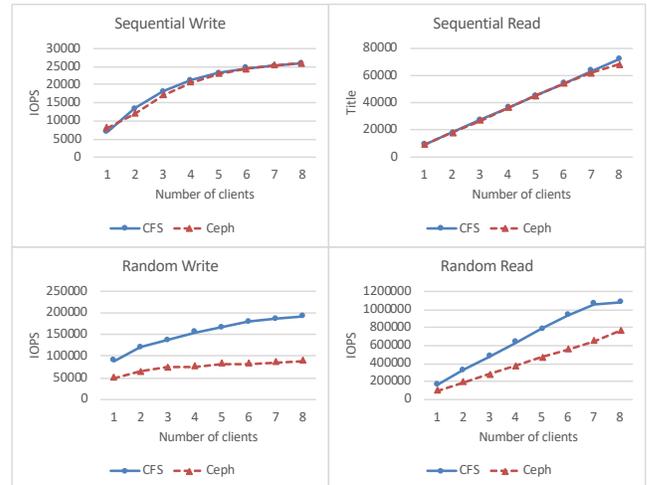}
\caption{IOPS when different number of clients. Each client has 64 processes in the random read/write tests and 16 processes in the sequential read/write tests. Each process operates a separate 40 GB file.}
\label{fig:multi}
\end{center}
\end{figure}

Next, we study how CFS and Ceph perform  in a multi-client environment with  the same set of tests. In this experiment,  each client has 64 processes in the random read/write tests and 16 processes in the sequential read/write tests, where each process operates a separate 40 GB file.  Each client in Ceph operates different file directories and each directory is bonded to a specific MDS in order to maximize the concurrency and improve the performance stability.  As can be seen in  Figure~\ref{fig:multi}, CFS has  significant performance advantage over Ceph  in the random read/write tests,   although their  performances in the sequential read/write tests are quite similar. These results are consistency with the ones in the previous single-client experiment, and can be explained in a similar way. 

To  sum up, in a highly concurrent environment,  CFS outperforms Ceph  in our random read/write tests for  large files.



\subsection{Small Files}

\begin{figure}[t!]
\begin{center}
\includegraphics[page=1,width=.48\textwidth]{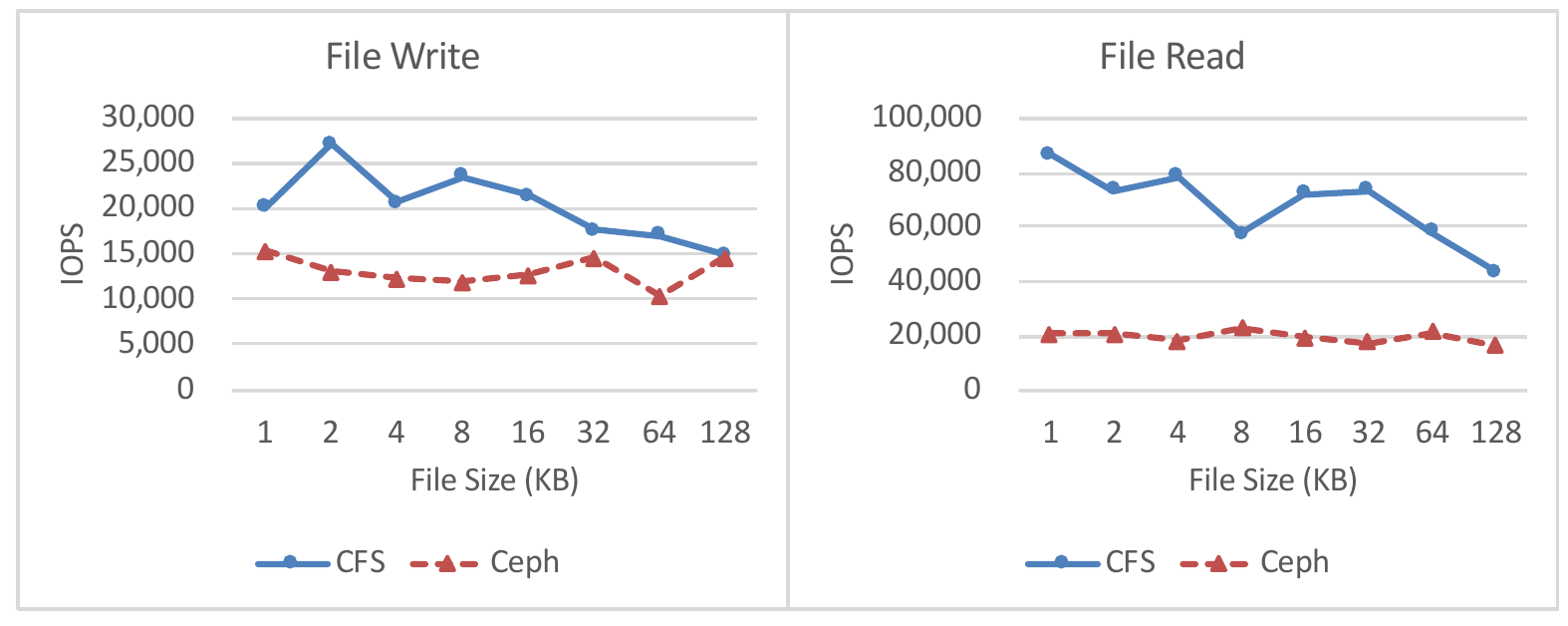}
\caption{IOPS when 8 clients, each of which has 64 processes, operate small files with different sizes.}
\label{fig:small}
\end{center}
\end{figure}

In this section we  study the performance of operating small files with various sizes  from 1 KB to 128 KB in CFS and Ceph. Similar to the metadata test, the results are obtained from \textit{mdtest}.  This experiment simulates the  use case when operating product images,  which are usually never modified once created. 
In our CFS configuration, 128 KB is the threshold set to determine if the file should be aggregated in a single extent or not, i.e., if we should treat it as  "small file" or not.  In Ceph, each client operates different file directories and each directory is bonded to a specific MDS in order to maximize the concurrency and improve the performance stability. 
 As can be seen from Figure~\ref{fig:small}, CFS outperforms Ceph in both read and write tests. This can be reasoned by the fact that,  (1)  CFS  stores all the file metadata in memory to avoid the expensive disk IOs during file read; and (2) in the case of small file write, the CFS client does not need to ask the resource manager for new extents; instead, it send the write request to the data node directly, which further reduces the network overheads. 

\section{Related Work}
\label{related}

GFS~\cite{gfs} and its open source implementation HDFS~\cite{hdfs} are designed for storing large files with sequential access. Both of them adopt the master-slave architecture~\cite{Thekkath}, where the single master stores all the file metadata. Unlike GFS and HDFS, CFS employs  a separate  metadata subsystem  to provide a scalable solution for   the  metadata storage so that the resource manager has less chance to become the bottleneck.  

Haystack~\cite{haystack}  takes after log-structured filesystems ~\cite{lsm} to serve long tail of requests seen by sharing photos in a large social network. The key insight is to avoid disk operations when accessing metadata.   CFS adopts similar ideas 
by putting the file metadata into the main memory. 
 However, different from haystack,  the actually physical offsets instead of logical indices of the file contents are stored in the memory, and deleting a  file  is achieved by the punch hole interface provided by the underlying file system instead of relying on the garbage collector to perform merging and compacting regularly for more efficient disk utilization. In addition, Haystack does not guarantee the strong consistency among the replicas when deleting the files, and it needs to perform merging and compacting regularly for more efficient disk utilization, which could be a performance killer.
Several works ~\cite{small1, small2} have proposed  to manage small files and metadata more efficiently by grouping related files and metadata together intelligently. CFS takes a different design principle to separate the storage of file metadata and contents. In this way, we can have more flexible and cost-effective deployments of meta and data nodes.


Windows Azure Storage (WAS)~\cite{azure} is a cloud storage system that provides strong consistency and multi-tenancy to the clients. Different from CFS, it builds an extra partition layer to handle random writes before streaming data into the lower level. AWS EFS~\cite{efs} is a cloud storage service  that provides scalable and elastic file storage. We could not evaluate the performance of Azure and AWS EFS comparing with CFS as they are not  open-sourced.

PolarFS~\cite{ali}  is a distributed file system designed for the Alibaba’s database service by utilizing  a lightweight network stack and I/O stack to take  advantage of the emerging techniques like RDMA, NVMe, and SPDK. OctopusFS~\cite{OctopusFS} is  a distributed file system based on HDFS  with automated data-driven policies for managing the placement and retrieval of data across the storage tiers of the cluster, such as memory, SSDs, HDDs, and remote storage.  These are orthogonal to our work as CFS can also adopt similar techniques to fully utilize the emerging hardware and storage hierarchies.
 
There are a few distributed file systems that have  been integrated with Kubernetes ~\cite{ceph, moosefs, gluster}. Ceph~\cite{ceph} is a petabyte-scale object/file store that maximizes the separation between data and metadata management by employing a pseudo-random data distribution function (CRUSH) designed for heterogeneous and dynamic clusters of unreliable object storage devices (OSDs). We have performed a comprehensive comparison with Ceph in this paper.
 GlusterFS~\cite{gluster} is a scalable distributed file system that aggregates disk storage resources from multiple servers into a single global namespace.  MooseFS~\cite{moosefs} is a fault- tolerant, highly available, POSIX-compliant, and scalable distributed file system. However, similar to HDFS, it employs a single master to manage the file metadata. 

MapR-FS\footnote{\url{https://mapr.com/products/mapr-fs/}} is a POSIX-compliant distributed file system that provides reliable, high performance, scalable, and full read/write data storage. Similar to Ceph, it stores the metadata in a distributed way alongside the data itself. 

\section{Conclusions and Future Work}
\label{conclude}
This paper describes CFS, a distributed file  system designed for serving JD's e-commence business.
CFS has a few unique features that differentiates itself from other open source solutions.  For example, it has a general-purpose storage engine with scenario-aware replications to accommodate different file access patterns with optimized performance.  In addition, it  employs a separate cluster to store the file metadata in a distributed fashion with a simple but efficient metadata placement strategy. Moreover, it provides POSIX-compliant APIs with relaxed  POSIX semantics and metadata atomicity to obtain better system performance.



We have implemented most of the operations by following the POSIX standard,  and are actively working on the remaining ones, such as  \textit{xattr,  fcntl, ioctl, mknod} and \textit{readdirplus}. 
In the future, we  plan to take advantage of the Linux page cache to speed up the file operations,  improve the file locking mechanism and cache coherency, and  support emerging hardware standards such as RDMA.  We also  plan to develop our own POSIX-compliant file system interface in the kernel space  to completely eliminate the overhead from FUSE.
%

\section{Acknowledgments}
We thank the anonymous reviewers for their valuable suggestions and opinions that helped improve the paper. We also express our
gratitude to Professor Enhong Chen of University of Science and Technology of China (USTC) who offered significant help for this work.

\bibliographystyle{ACM-Reference-Format}
\balance
\bibliography{citation}


\end{document}